\documentclass{article}
\usepackage[utf8]{inputenc}
\usepackage[T1]{fontenc}
\usepackage{hyperref}
\usepackage{url}
\usepackage{amsmath}
\usepackage{graphicx}
\usepackage{authblk}
\usepackage{algorithm}
\usepackage{algpseudocode}

\usepackage{caption}
\usepackage{subcaption}
\usepackage[backend=biber, style=ieee]{biblatex}
\usepackage{geometry}
\usepackage{float}
\usepackage{comment}
\usepackage{multirow} 

\begin{document}

\title{MARVEL: A Multi Agent-based Research Validator and Enabler using Large Language Models}

\author[1,2]{Nikhil Mukund}
\author[3]{Yifang Luo}
\author[1,3]{Fan Zhang}
\author[1,2]{Lisa Barsotti}
\author[1]{Erik Katsavounidis}

\affil[1]{MIT Kavli Institute for Astrophysics and Space Research and LIGO Laboratory,\\
    Massachusetts Institute of Technology, Cambridge, MA 02139, USA}
\affil[2]{NSF AI Institute for Artificial Intelligence and Fundamental Interactions (IAIFI), Cambridge, MA, USA}
\affil[3]{State Key Laboratory of Ocean Sensing \& Ocean College, Zhejiang University, Zhoushan, 316021, China}

\maketitle

\begin{abstract}
We present MARVEL (\url{https://ligogpt.mit.edu/marvel}), a locally deployable, open-source framework for domain-aware question answering and assisted scientific research. It is designed to address the increasing demands of a digital assistant for scientific groups that can read highly technical data, cite precisely, and operate within authenticated networks. 
MARVEL combines a fast path for straightforward queries with a more deliberate DeepSearch mode that integrates retrieval-augmented generation and Monte Carlo Tree Search. It explores complementary subqueries, allocates more compute to promising branches, and maintains a global evidence ledger that preserves sources during drafting. We applied this framework in the context of gravitational-wave research related to the Laser Interferometer Gravitational-wave Observatory. Answers are grounded in a curated semantic index of research literature, doctoral theses, LIGO documents, and long-running detector electronic logbooks, with targeted web searches when appropriate. Because direct benchmarking against commercial LLMs cannot be performed on private data, we evaluated MARVEL on two publicly available surrogate datasets that capture comparable semantic and technical characteristics. On these benchmarks, MARVEL matches a GPT-4o mini baseline on literature-centric queries and substantially outperforms it on detector-operations content, where domain retrieval and guided reasoning are decisive. By making the complete framework and evaluation datasets openly available, we aim to provide a reproducible foundation for developing domain-specific scientific assistants. 
\end{abstract}

\section{Introduction}

Large scientific collaborations generate substantial technical materials, including papers, reports, software, wikis, meeting notes, and internal discussions. As projects progress and team members change roles, a lot of this knowledge becomes scattered. When key details are hard to locate, earlier work is sometimes repeated. Early-career researchers often struggle to navigate extensive collections of past documents, and finding relevant prior work can be time-consuming. To reduce these inefficiencies, research groups need improved methods for accessing existing information. Retrieval and recommendation tools can play a crucial role in ensuring that knowledge is reused rather than reinvented. These tools have evolved from simple keyword searches to advanced research assistants capable of supporting complex reasoning and informed decision-making. Such tools can also help preliminarily validate new research ideas by finding supporting evidence from existing sources. Relying primarily on commercial frameworks that use closed-source large language models (LLMs) limits the ability to optimize such systems based on collaboration requirements, such as prioritizing specific datasets, using specific LLMs, or allocating task-specific compute resources. The specifics of the algorithms used in tasks such as reasoning are often obscured, limiting our ability to optimize them further. Tools that perform intricate multi-step reasoning often require hundreds of LLM inference calls, which increases overall cost.

We present MARVEL, a locally deployable, open-source framework for domain-specific question answering and research assistance. In this study, we have applied MARVEL to gravitational-wave research, using the Laser Interferometric Gravitational-Wave Observatory (LIGO) as the main test case. We use domain-aware retrieval with a compute-aware exploration module that integrates and verifies information from multiple sources before synthesis, enabling reliable, citation-grounded answers. We make use of open-weight LLMs for reasoning and structured task execution, and formulate novel search strategies over sources spanning arXiv papers, curated theses, technical documents, and detector electronic logbooks. The framework provides a transparent and reproducible architecture for retrieval, re-ranking, search allocation, and report generation. Although demonstrated here for topics related to LIGO and gravitational-wave instrumentation, extension to other scientific domains requires only swapping datasets and making minor prompt adjustments to fit the new context.

The paper is organized as follows. Section \ref{sec:context} describes the setting of gravitational-wave observatories and their information handling practices. Section \ref{sec:related} reviews related work and document characteristics. Section \ref{sec:methods} presents the technical approach, including the DeepSearch algorithm. Section \ref{sec:evaluation} reports the evaluation results. Section \ref{sec:conclusion} offers concluding remarks and possible directions for further work.

\section{Gravitational-wave observatories and information handling}
\label{sec:context}

About a hundred years after Einstein predicted the existence of gravitational waves (GWs), their first direct detection in 2015 confirmed a key prediction of general relativity and added a new way to study the universe. Within less than a decade, the field has progressed to regular detections of compact-object mergers, showing a shift from discovery to precision measurements. This progress has been made possible by a global network of kilometer-scale interferometers, which includes LIGO in the United States~\cite{TheLIGOScientificCollaboration2015}, Virgo in Europe~\cite{acernese2015}, KAGRA in Japan~\cite{aso2013}, and GEO600 in Germany~\cite{dooley2016}. Detectors like including LIGO-India~\cite{Unnikrishnan2013}, Cosmic Explorer~\cite{reitze2019cosmic}, the Einstein Telescope~\cite{punturo2010}, and the space-based LISA mission~\cite{amaro2017}, is likely to begin observations within the coming decade. These instruments will expand the network and increase its sensitivity, enabling scientists to observe gravitational waves over a wider frequency range and from more distant sources.

The detection of GW170817~\cite{Abbott2017GW170817}, the first binary neutron star merger observed by the Advanced LIGO and Advanced Virgo detectors, was also followed up by telescopes operating in optical, X-ray, and gamma-ray regimes. Since then, coordinated efforts among many observatories have enabled the rapid location and study of astrophysical events within minutes of a detection alert. As the number of detectors and partner observatories continues to grow, the supporting information systems have also become larger and more complex. Improvements in detector sensitivity and analysis algorithms have led to steady progress in the number and quality of event detections. So far, the worldwide network has recorded more than 300 gravitational-wave events from sources such as merging black holes and neutron stars, and we expect the number to rise with planned future upgrades. This scientific growth is supported by strong systems engineering and project management practices that keep detectors in good condition, organize observing runs, and improve collaboration between laboratories.

Recent observing campaigns have produced catalogs of compact binary mergers~\cite{GWTC1_2019,GWTC21_2021,GWTC3_2021},
with associated open data releases through the Gravitational Wave Open Science Center~\cite{Abbott2021O1O2OpenData,GWOSCData}.
The main data-analysis software stacks used within the LVK for these searches include the LIGO Algorithm Library Suite (LALSuite)~\cite{LALSuite2020},
the PyCBC pipeline~\cite{Usman2016PyCBC}, and the GstLAL search framework~\cite{Cannon2021GstLAL}.

 Each detector is a complex optomechanical system comprising high-power lasers, vibration-isolation platforms, suspended optics, hundreds of sensors and actuators, and multiple digital and analog control loops. Operating these interferometers is a sustained systems-engineering effort. Upgrades and commissioning alternate with long observing campaigns, during which noise mitigation, calibration, and data-quality assessment must continue with minimal intrusion into astrophysical data taking. The collaboration also develops and maintains several software stacks for signal detection, parameter estimation, calibration, and detector diagnostics. These codebases have been built over many years and continually refined as the instruments evolve. The knowledge required to develop, use, and maintain these systems is, however, dispersed across decades of electronic logbooks, document databases, internal wikis, meeting minutes, issue trackers, and published literature. Heterogeneous formats, evolving terminology, and the fact that people move between projects over time make manual search slow and difficult. Practical problems at the detector sites tend to recur over periods of several months to a few years, and solutions developed at one site are often helpful at the others. A system that can gather this scattered technical knowledge and respond to queries would therefore be valuable for day-to-day work across the observatories.

\section{Related Work and Data Characterization}
\label{sec:related}

\subsection{Prior work: HeyLIGO}
An early effort in this direction was the HeyLIGO service (\url{https://heyligo.gw.iucaa.in}), an information-retrieval and recommendation tool based on natural-language processing (NLP). It was launched in 2017 to index the electronic logbooks from the LIGO Hanford, LIGO Livingston, GEO\,600, and Virgo observatories, and it provided a web interface for semantic search and exploration~\cite{mukund2018information}. By quickly displaying relevant logbook entries and associated images, HeyLIGO demonstrated how NLP can support day-to-day detector operations. It relies on classic information-retrieval methods based on TF-IDF~\cite{salton1986introduction,salton1988term} together with Word2Vec embeddings~\cite{mikolov2013efficient,mikolov2013distributed} to search and relate content within the detector logbooks. As new logbook entries are added, HeyLIGO periodically reprocesses the corpus and retrains its embedding model so that the vector space reflects the evolving vocabulary and maintains retrieval quality~\cite{mukund2018information}. But the service was primarily built for a single document type and relied on a single retrieval method. It lacks the LLM-based capabilities that have become available since then, as well as the cross-document reasoning needed to provide context-based answers. In this work, we use HeyLIGO as one of the logbook-focused retrievers as part of the hybrid search strategy, contributing candidate entries that are combined with other search results.

\subsection{Semantic Clustering of Documents}

\begin{figure}[H]
\centering

\begin{subfigure}[t]{0.48\textwidth}
    \centering
    \includegraphics[width=\textwidth]{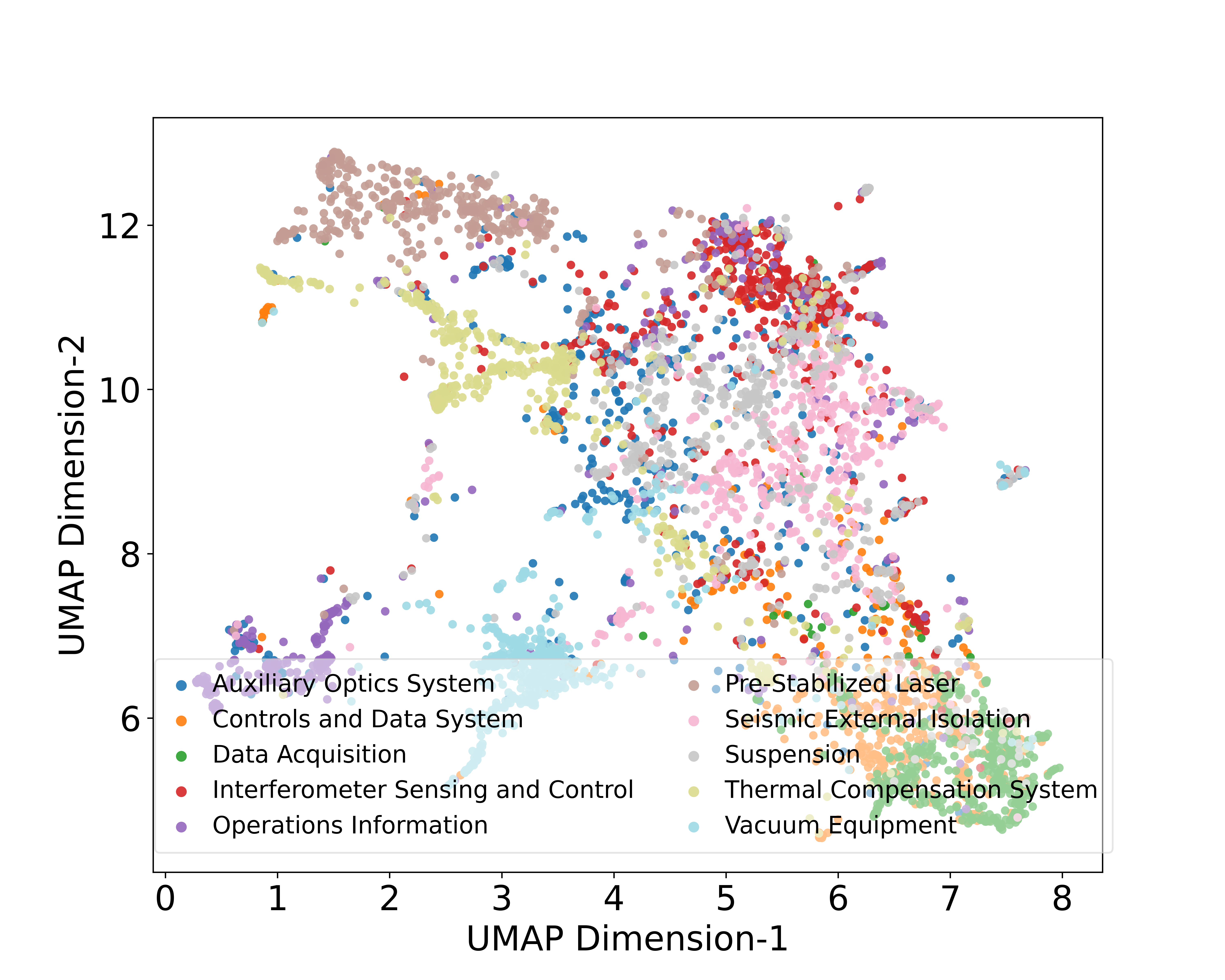}
    \caption{LIGO electronic logbooks}
    \label{fig:logbook-umap}
\end{subfigure}
\begin{subfigure}[t]{0.48\textwidth}
    \centering
    \includegraphics[width=\textwidth]{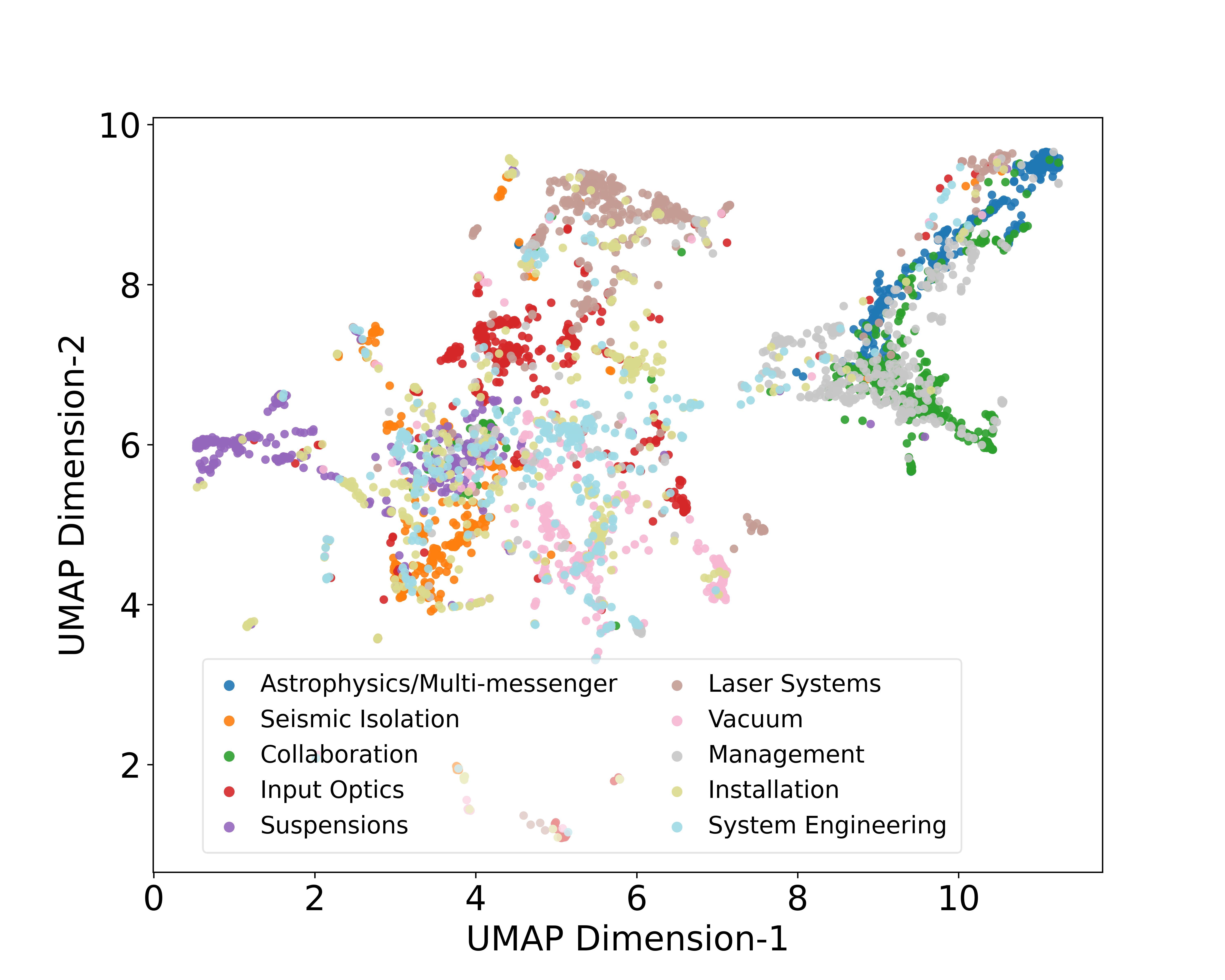}
    \caption{LIGO technical documents}
    \label{fig:dcc-umap}
\end{subfigure}

\caption{UMAP projections of dense text embeddings, using 500 documents per category from the ten most frequent classes in each text corpus.  
(a) Logbook entries contain topics related to routine operational activity at the detector sites.  
(b) Technical documents cover topics related to LIGO and broader gravitational-wave related fields.}
\label{fig:embedding-landscape-comparison}
\end{figure}

To examine how the documents relate to each other, we generated a two-dimensional projection of the embeddings for the ten most common topics in the LIGO logbooks~\cite{LIGOaLOGHanford,LIGOaLOGLivingston} and the LIGO Document Control Center (DCC)~\cite{LIGODCC}, as shown in Figure \ref{fig:embedding-landscape-comparison}. The embedding procedure is described in Section \ref{subsec:corpus_embed}, and UMAP~\cite{mcinnes2018umap} was used for the dimensionality reduction. Documents linked to the same topic form clear clusters, while topics with closely related technical content appear near one another in the projection. This projection also shows the limits of relying only on semantic similarity. Many detector terms, channel names and acronyms carry very specific meanings that may not be captured well by an embedding. For these cases, a straightforward keyword search is often more reliable. However, purely lexical matches can also pull in material that uses the exact wording but is not actually relevant.
For these reasons, we adopt a hybrid retrieval scheme that combines semantic and keyword-based search, followed by re-ranking and a verification step, as described in the following sections.

\section{Technical Approach and Methodology}
\label{sec:methods}
\begin{figure}[H]
\centering
\includegraphics[width=\textwidth]{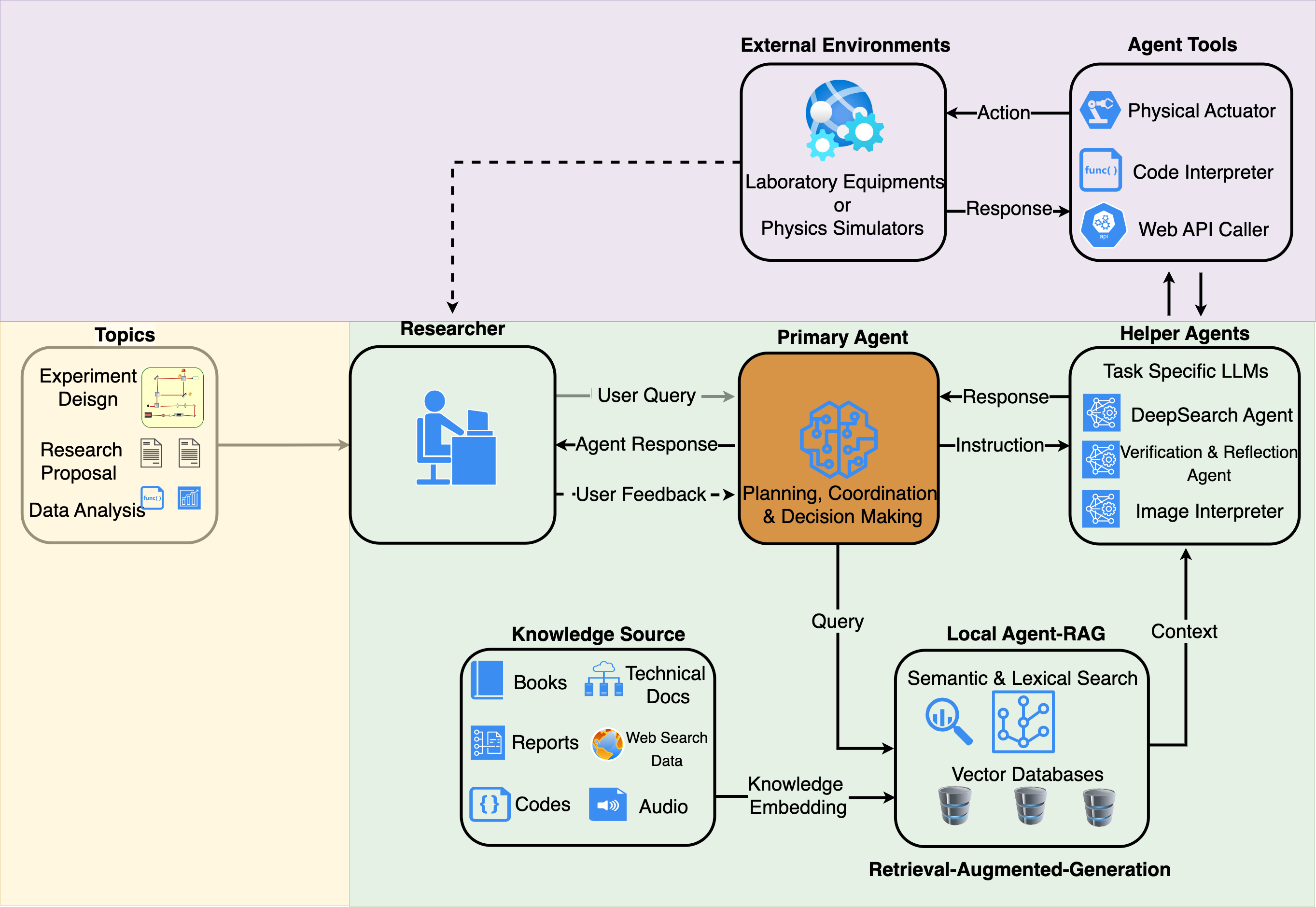}
\caption{
Overview of the MARVEL architecture and the status of the current implementation. 
The central agent manages the helper modules, the available tools, and the retrieval system built from the document embeddings. 
The green box shows the parts implemented in this work. 
Yellow box indicates the research tasks that MARVEL aims to support. Future work is marked in purple.
}
\label{fig:MARVEL}
\end{figure}

\begin{figure}[H] \centering \includegraphics[width=0.8\linewidth]{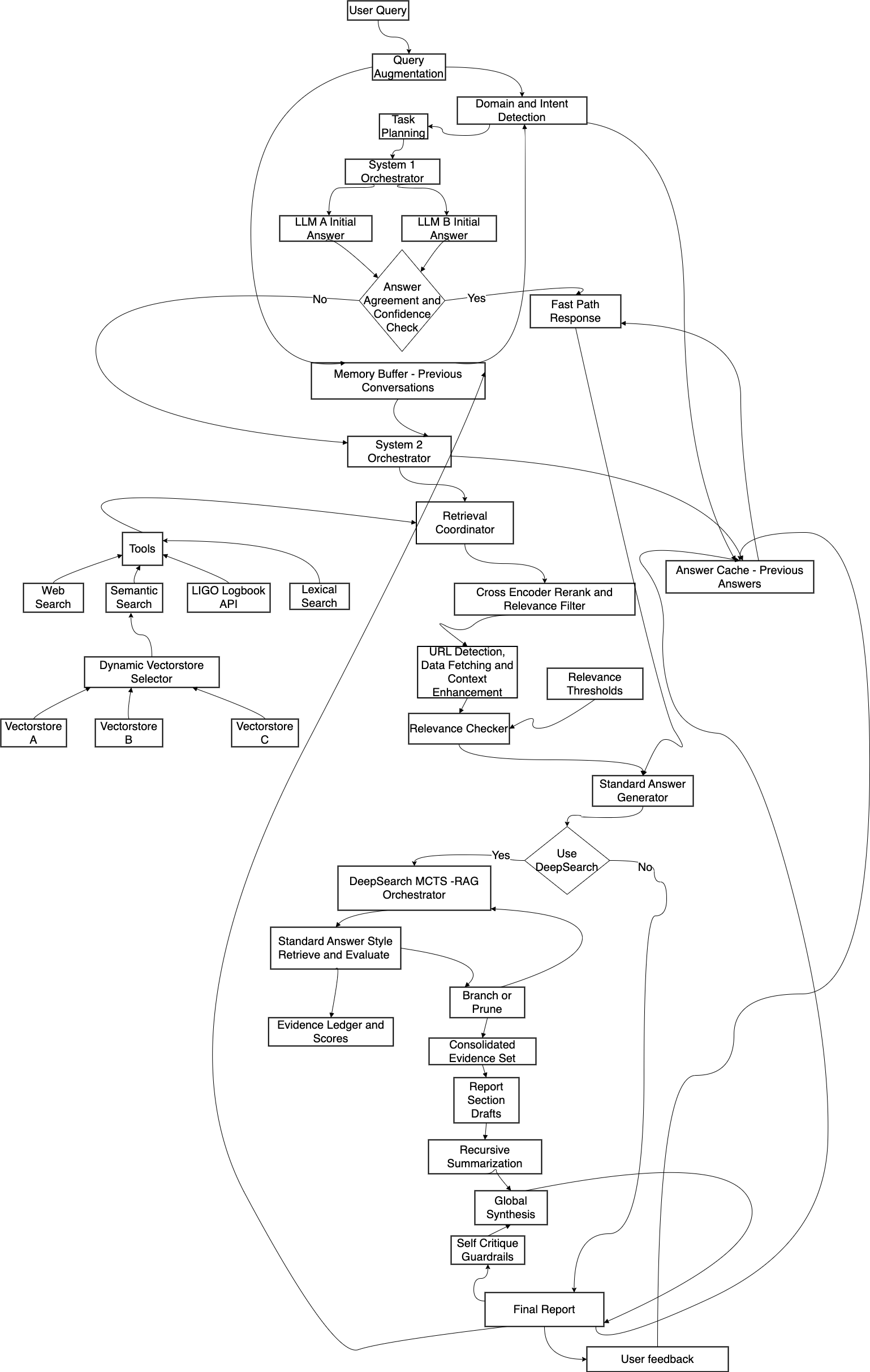} \caption{MARVEL workflow. A user query is first processed by the primary agent for task planning, domain detection, and initial reasoning (System-1). If agreement between fast-path answers is low, the query escalates to the System-2 orchestrator for retrieval-augmented generation (RAG) using hybrid semantic, lexical, and web-based searches. Relevant contexts are re-ranked, verified, and merged before synthesis into a citation-grounded draft. When necessary, the DeepSearch controller invokes a Monte Carlo Tree Search (MCTS) based exploration to collect complementary information, allowing the system to generate a cited report even when the evidence is spread across multiple sources.} \label{fig:MARVEL-detailed} \end{figure}

Figure~\ref{fig:MARVEL} illustrates the MARVEL setup. A central agent coordinates the main workflow, while helper agents execute individual tasks through tool interfaces.
Tasks scheduled for future development, including limited interaction with laboratory hardware or simulators, are highlighted in purple.
Retrieval is performed via the local retrieval-augmented generation (RAG) system built from document embeddings.
The colored boxes indicate the parts implemented in this work and the areas planned for later development.
Figure~\ref{fig:MARVEL-detailed} then outlines in detail how MARVEL processes a user query. When a query is submitted, the system requests clarifications when needed through a short interaction implemented using a finite-state machine. Based on the inferred intent, the planning agent selects the required data sources and utility tools for retrieval, re-ranking, verification, and report generation. The helper agents and the retrieval-augmented generation work together to produce a cited answer. The main subcomponents of MARVEL are described in the following sections.

\subsection{Background: LLMs, retrieval, and agents}
LLMs are auto-regressive text token predictors that generate the next word in a sequence conditioned on a given prompt. They use the transformer architecture~\cite{vaswani2017attention}, which allows them to compress extensive collections of textual data within their model weights. However, there are a few limitations when such foundational LLMs are directly used as scientific assistants. The first is that most of the model's knowledge is stored implicitly in its parameters, which makes it hard to update, verify, or connect to a clear source of reference. The second is the limited-size context window, akin to working memory, which limits both the amount of text that can be processed at once and the level of attention it can provide to individual sections within that text. This makes it difficult to reason over long documents or data from several sources. Approaches such as Retrieval-Augmented Generation (RAG)~\cite{lewis2020rag} can mitigate these issues to some extent. RAG adds a retrieval step at query time to bring in relevant information from external databases, improving factual accuracy and transparency. The documents are divided into smaller sections, converted into dense vector embeddings, and stored in a searchable index. The user's query is also converted to the same vector space, and the most similar text sections are retrieved using both semantic and lexical matching. These selected data chunks are then provided to a locally hosted LLM as context, which then produces a response supported by those sources. While vanilla RAG improves factual correctness and recall compared to models that rely solely on their internal memory, the additional steps described in this section further enhance the quality of the generated answer.

The capability of an LLM to reason, self-critique, and plan is also limited unless used in conjunction with external tools and algorithms that provide it with agentic capabilities. In reinforcement learning, an agent is usually described as a system that can observe its surroundings, make decisions, take actions, and learn from the resulting experiences~\cite{sutton2022intelligent}. In this work, we use the term agent to refer to a system that emulates this idea, with an LLM serving as the primary reasoning component and having access to tools for executing functions and managing memory. The agent, however, does not have complete autonomy; instead, its reasoning and tool-use abilities are achieved through an orchestration layer depicted in Figure ~\ref{fig:MARVEL-detailed}, with appropriate prompts tuned for task completion.

\subsection{LLM selection and deployment}
\label{sec:llm-selection}
For local deployment, MARVEL is designed to run on consumer-grade GPUs such as NVIDIA RTX-class hardware, which we used during development and testing. Models with 14-30 billion parameters and 4-bit quantization offer a good balance between accuracy and response time while remaining practical for such systems. Sparse mixture-of-experts models are particularly efficient because only a small part of the network is active during inference~\cite{jiang2024mixtral}. A long context window (128k tokens or more) is also essential for combining information from multiple sources.
For the primary agent, we use the \texttt{meta-llama/llama-4-scout-17b-16e-instruct} model~\cite{llama4scoutHF}, which follows a mixture-of-experts design with about 17 billion active parameters. Local inference is carried out using the Ollama server~\cite{ollamaDocs}, which supports quantized models, concurrent serving, and easy migration to newer open-weight models as they are released. During evaluation and prototyping, we used both local GPU setups and Groq's LPU backend~\cite{groq2024lpu}. For the public demonstration, some LLM calls are routed through Groq's backend to provide faster responses for many concurrent users. The intended private and research deployments of MARVEL run entirely on local hardware.

\subsection{Corpus construction and semantic indexing}
\label{subsec:corpus_embed}
To support semantic search, we created a text corpus covering the last ten years of arXiv submissions containing the keyword "LIGO," curated Ph.D.\ theses, about 14,000 publicly accessible LIGO documents, and years of public LIGO detector logbook entries. PDF files were ingested and processed with optical character recognition (OCR) and layout recovery to preserve mathematics, tables, and figure captions. This processing was carried out on the MIT Engage cluster using a pair of H200 GPUs. We used Surya~\cite{paruchuri2025surya} for OCR with layout detection, and Marker~\cite{paruchuri2025marker} for PDF-to-markdown conversion that retains document structure. The corpus was normalized and deduplicated by removing boilerplate text and near-duplicates using a high cosine-similarity threshold between embeddings. We then split each document into 1500-token chunks with a 500-token overlap, and the chunks inherited parent metadata whenever available. For embeddings, we used \texttt{nomic-embed-text}, an open long-context encoder (8K tokens) selected for its reproducible retrieval performance, low latency, and permissive licensing~\cite{nomic2023embed}.

\subsection{Retrieval and re-ranking}
At query time, MARVEL performs a hybrid retrieval that combines regular keyword matching with dense vector search over the semantic index. For lexical search, we use BM25+~\cite{lv2011bm25plus}, which improves on classical BM25~\cite{robertson1995okapi} by handling document-length variation and reducing the bias toward shorter documents. A key component that enables semantic retrieval at scale is the vector database. When datasets become very large, direct or brute-force embedding search is too slow for real-time use. The FAISS (Facebook AI Similarity Search) library~\cite{johnson2019billion,khanda_agentic_2024} provides algorithms for approximate nearest-neighbor search and clustering of dense vectors, giving millisecond-level response even for millions of entries. By using FAISS, MARVEL can quickly retrieve semantically relevant passages while keeping the system responsive. The retrieved passages are then re-ranked using ColBERTv2~\cite{khattab2020colbert,santhanam2022colbertv2}, which improves token-level alignment of technical terms without the high cost of a full cross-encoder. Metadata linked to each retrieved item is used to collect citation details, source URLs, publication dates, and other bibliographic information when available. Web-based results are also collected using the Tavily API~\cite{tavily2024api}, a publicly available web search service that provides structured outputs containing titles, URLs, short content summaries, and relevance scores, making them easy to use within our retrieval pipeline.

\subsection{Answer-aware chunk filtering and generation}
\label{sec:chunk-verification}

When too many retrieved chunks are combined, even after re-ranking, the model often becomes confused, and the quality of the answer declines. LLMs do not attend reliably to very long inputs and may miss useful information in the middle of the context~\cite{liu-etal-2024-lost}. Adding extra passages can also introduce irrelevant text, reducing performance in multi-document tasks~\cite{wang2024corag}. To handle this, we use a simple chunk-verification step between retrieval and generation. For each of the top-$k$ passages, a smaller LLM checks whether the text likely contains evidence related to the query. If it does, only the relevant portions are kept. Passages that are not relevant are removed. The remaining text is shortened, cleaned for duplicates, and stored with its source tags to keep citation accuracy. The main generator then operates on this smaller, focused context rather than the complete set of retrieved documents. Although this introduces a slight delay, it improves both the relevance and correctness of the answers, consistent with earlier studies on answer-aware filtering in RAG systems~\cite{asai2023selfrag}.

\subsection{DeepSearch via MCTS-augmented RAG}
\label{sec:deepsearch}
MARVEL-Standard search is often sufficient for well-posed, straightforward questions, but it can fall short on technical questions that need deeper reasoning. Such questions usually require asking several focused sub-questions, checking their answers, and planning the next steps based on the evidence gathered. Relying only on the initial query can bottleneck the entire pipeline. If the first retrieval misses important documents or interpretations, every downstream step inherits that gap. We already use rephrasing, acronym expansion, and query augmentation in the standard search path to reduce this risk, but these remain single-hop heuristics~\cite{Lewis2020RAG}. Two challenges dominate deep exploration. First, the search space is combinatorial, and the number of useful sub-questions grows quickly with topic breadth. Second, language models are sensitive to retrieval quality, and without guidance, they may pursue weak branches or collapse to local optima. Contemporary prompting strategies such as self-ask, ReAct, tree-of-thoughts, and reflection partially address these issues by breaking problems into steps and using feedback~\cite{Press2022SelfAsk, Yao2022ReAct, Yao2023ToT, Shinn2023Reflexion}. However, they still require a principled way to balance exploration and exploitation when selecting which sub-questions to follow, and they benefit from an explicit measure of novelty in the evidence. A more reliable approach is to allocate additional test-time compute~\cite{Wang2022SelfConsistency} to examine multiple aspects of the question, expand a search frontier, and synthesize the final answer from the union of high-value evidence. This is the aim of DeepSearch. It combines a standard single-query RAG system with a Monte Carlo Tree Search (MCTS)~\cite{Kocsis2006UCT, Browne2012MCTS, Silver2016AlphaGo} controller that allocates more computation to sub-queries that appear most promising at test time. Recent studies have also used MCTS together with retrieval for multi-step reasoning~\cite{tran2024rare,lee2024mzqa,hu2025mctsrag,chen2025sic}. Our setup differs in a few practical ways. We use a two-factor score that checks both the answer quality and the relevance of each sub-query to the main question. This keeps the search aligned with the original query. We also maintain a global evidence ledger with stable inline markers, apply global duplicate control, use a simple early-stop rule, and carry all citations through a hierarchical synthesis stage. These features are explained below.

\subsubsection{Formulation}
DeepSearch approaches multi-hop scientific question answering as a guided exploration task. It combines a careful single-query RAG system with an MCTS controller and a source-tracking synthesis module to generate answers that are richer in content and better supported by citations than those from a single pass. At the same time, the method avoids unnecessary computation through reflection, duplicate control, and early stopping. To keep track of information across the search tree, DeepSearch uses a global evidence ledger. This ledger is a structured index that stores each retrieved document along with its identifier, such as a DOI, arXiv ID, LIGO document ID, or normalized URL. It also records basic metadata, such as the title, source, and publication date. This helps the system identify new information, reduce duplication, and maintain clear links between citations and their supporting evidence during multi-hop reasoning.

Algorithm~\ref{alg:deepsearch} outlines the key steps used to implement DeepSearch. We briefly describe its components here. Let $q_0$ be the user query and $\mathcal{C}$ the document collection. A MARVEL-Standard search returns an initial answer $a(q_0)$ together with the supporting documents $\mathcal{D}(q_0)\subset\mathcal{C}$. DeepSearch builds a search tree $T$ whose nodes are sub-queries ${q_i}$ derived from $q_0$. Each node stores the sub-query, the corresponding answer text produced by the retriever, the retrieved documents, a scalar score, and links to child nodes that explore deeper aspects of the question. If $A(q)$ denotes the answer text and $\text{ctx}(q)$ the path context, the score is

\begin{equation}
\label{eq:node-score-corrected}
s(q) = \mathrm{EvalAnswer}(A(q), q)\; \times\; \mathrm{EvalQuery}(q, \mathrm{ctx}(q)).
\end{equation}

The first component of the product evaluates the quality of the sub-query answer, while the second examines the relevance of the sub-query to the original query. This product acts as an approximate reward signal for MCTS. We choose the next child node during selection using the Upper Confidence Bound for Trees (UCT) rule~\cite{Kocsis2006UCT}:

\begin{equation}
\label{eq:uct}
\mathrm{UCT}(n) = \overline{X}_n + c\cdot\sqrt{\frac{\ln N_p}{N_n}} ,
\end{equation}

where $\overline{X}_n$ is the running average of backpropagated rewards at node $n$,
$N_p$ and $N_n$ are the visit counts of the parent and child, and $c>0$ sets the exploration level.
After each simulation, the reward $s(q)$ from the selected child is backpropagated along the path,
updating both the visit counts and the value sums that determine $\overline{X}_n$. New sub-queries are generated through an LLM-based prompt conditioned on the path context $(q_0, a(q),\mathcal{D}(q))$. To prevent repetitive exploration, each proposed sub-query is compared with a global set of previously used sub-queries via their embedding vectors, and near-duplicates are discarded. The node is then expanded by retrieving documents for each accepted sub-query and computing its score using Eq.~\eqref{eq:node-score-corrected}.
Although the search runs for a fixed maximum number of iterations, it is terminated if the best cumulative path score fails to improve by more than a small threshold for several successive iterations. Our implementation keeps the single-query retriever unchanged and wraps it in an MCTS loop that generates only a small number of children per node, constrained by the available compute budget. Each retrieved chunk is assigned a stable inline marker, preserved through the drafting stage, and renumbered consistently at the end, which ensures that citations remain traceable even if the LLM reorders text.
Compared with fixed-depth reasoning or single-hop prompting, MCTS provides a clear exploration-exploitation trade-off through Eq.~\eqref{eq:uct}. It also allows reward signals from multiple sub-queries to be aggregated through backpropagation. In practice, small branching factors (four to five) and moderate values of $c(\approx1.4)$ provide good coverage at reasonable test-time cost. Our goal is to explore different reasoning paths, examine missing connections, and recover supporting evidence across multiple sub-queries.

\begin{algorithm}[H]
\caption{MARVEL-DeepSearch}
\label{alg:deepsearch}
\begin{algorithmic}[1]

\Require user query $q_0$; retrieval function Retrieve; sub-query generator GenSub;
         evaluators EvalAnswer and EvalQuery; iteration budget $I$; per-node expansion
         budget $B$; UCT constant $c$; duplicate threshold $\tau_{\mathrm{dup}}$;
         early-stopping threshold $\varepsilon$; patience $K$

\Ensure final cited report and diagnostics

\State $(a_0, \mathcal{D}_0) \gets \text{Retrieve}(q_0)$
\State create root node $n_0 \gets (q_0, a_0, \mathcal{D}_0)$
\State initialise visit counts and value sums for all nodes to zero
\State $E \gets \{\text{emb}(q_0)\}$ \Comment{global list of sub-query embeddings}
\State Expand($n_0, B, E, \tau_{\mathrm{dup}}$)

\State $B_{\text{prev}} \gets 0$;\quad $no\_improve \gets 0$

\For{$t = 1$ to $I$}

  \State \textbf{Selection:} starting at $n_0$, select a node $n$ by repeated UCT selection (Eq.~\ref{eq:uct})
         until reaching a node that can be expanded or has unvisited children

  \If{$n$ is not expanded}
     \State Expand($n, B, E, \tau_{\mathrm{dup}}$)
  \EndIf

  \If{$n.\text{children} = \emptyset$}
     \State \textbf{break} \Comment{no further exploration possible}
  \EndIf

  \State \textbf{Simulation:} define $U = \{ ch \in n.\text{children} \mid ch.\text{visits} = 0 \}$
  \State choose child $m$ from $U$ if $U \neq \emptyset$, otherwise from $n.\text{children}$
  \State $r \gets m.\text{score}$

  \State \textbf{Backpropagation:} update visit counts and value sums for all nodes on the path from $n_0$ to $m$

  \State compute best root-to-leaf cumulative score $B_t$

  \If{$t>1$ and $B_t - B_{\text{prev}} < \varepsilon$}
     \State $no\_improve \gets no\_improve + 1$
  \Else
     \State $no\_improve \gets 0$
  \EndIf

  \If{$no\_improve \ge K$}
     \State \textbf{break}
  \EndIf

  \State $B_{\text{prev}} \gets B_t$

\EndFor

\State \textbf{Synthesis:}
\Statex \hspace{1.5em} $\triangleright$ enumerate root-to-leaf paths and rank them by cumulative score
\Statex \hspace{1.5em} $\triangleright$ generate updated answers relative to the baseline
\Statex \hspace{1.5em} $\triangleright$ extract citation URLs and attach stable inline markers
\Statex \hspace{1.5em} $\triangleright$ run the LangGraph pipeline to produce \textsc{Key Facts}, \textsc{Finer Details}, and \textsc{Summary}
\Statex \hspace{1.5em} $\triangleright$ build \textsc{References} from the unique collected URLs
\State \Return final report and diagnostics

\Statex \vspace{0.3em}
\Statex \rule{\linewidth}{0.35pt}
\Statex \vspace{0.15em}

\Function{Expand}{$n, B, E, \tau_{\mathrm{dup}}$}
   \State build context string from the path from $n_0$ to $n$
   \State attempts $\gets 0$
   \While{$|n.\text{children}| < B$ and attempts $< 4B$}
        \State attempts $\gets$ attempts $+ 1$
        \State $q' \gets \text{GenSub}(context, \text{existing siblings}, q_0)$
        \If{$\max_{e \in E} \langle \text{emb}(q'), e \rangle \ge \tau_{\mathrm{dup}}$}
            \State \textbf{continue} \Comment{global near-duplicate}
        \EndIf
        \State append $\text{emb}(q')$ to $E$
        \State $(a', \mathcal{D}') \gets \text{Retrieve}(q')$
        \State form answer text $A'$ from QA and summary fields
        \State $r' \gets \text{EvalAnswer}(A', q') \times \text{EvalQuery}(q', context)$
        \State add child $(q', A', \mathcal{D}')$ with score $r'$ to $n.\text{children}$
   \EndWhile
   \State mark $n$ as expanded
\EndFunction

\end{algorithmic}
\end{algorithm}

Even with long context windows, LLMs often fail to exploit the available information. Content placed near the middle of a prompt tends to receive less attention~\cite{Liu2023LostMiddle}. Asking the LLM to join the sub-answers and the retrieved documents with a single prompt often doesn't fit within the context window and also results in information loss. To handle this, we use a  task-specific synthesis built with a LangGraph pipeline~\cite{langgraph2024}. High-value nodes identified during the search are first summarized into short notes that keep their inline citations. These notes are then processed one after another through an initial drafting stage followed by iterative refinement. During this stage, new evidence is merged into a growing draft under defined task categories such as \textsc{Key Facts}, \textsc{Finer Details}, and \textsc{Summary}. These categories are intended to represent progressively broader levels of synthesis. The concluding stage reorders citations by first appearance and produces a deterministic reference list. This hierarchical integration allows ample evidence sets to be combined without exceeding token limits while maintaining a transparent and verifiable link between generated statements and their supporting sources. Because we use LLM as a judge, the value proxy $s(q)$ may overweight easy-to-measure lexical or embedding similarity and underweight subtle but consequential evidence. Future work will address this using learned critics or estimates of retrieval-time uncertainty. Although MCTS allocates computation adaptively, some budget can still be wasted on unproductive branches when the sub-queries drift away from the primary question. The evidence ledger is intended to prevent citation loss, but near-duplicate records still introduce redundancy. Strengthening the deduplication layer, along with novelty-promoting checks, improved sub-query proposals, and adaptive widening policies, can improve the overall search behavior and help keep the evidence base compact and easier to verify.

\section{Evaluation}
\label{sec:evaluation}

Evaluating scientific research assistants requires both quantitative metrics and qualitative checks to determine whether responses are accurate, traceable, and valuable in practice. We use three complementary evaluation procedures that together provide a balanced measure of overall system performance. The experiments are carried out on two representative open datasets: ArXivData, which focuses on literature-style text and contains roughly 900 question-answer pairs, and LogbookData, which focuses on detector operations and contains about 700 pairs. GPT-4o mini serves as the commercial baseline for comparison. For each dataset, the evaluation corpus is created by generating question-answer pairs anchored to individual text chunks. These pairs are first produced automatically using GPT-4o mini and then manually inspected to ensure that each question is self-contained, clearly phrased, and has an objective, verifiable answer. This filtering removes vague, trivial, or overly open-ended cases and keeps the benchmark aligned with the intended use of scientific assistants. 

The first evaluation method uses GPT-4o mini as an impartial blind grader. For every question, the grader receives the context, the ground-truth answer, and two anonymized system responses in random order. It scores on a 0-1 scale for relevance and factual correctness relative to both the question and the provided context, with higher scores awarded for responses that cite evidence and a score of 0 assigned to responses that state an inability to answer. This emulates a domain-fluent judge scoring competing responses without access to system identity. The second evaluation applies the RAGAS framework (Retrieval-Augmented Generation Assessment Suite) framework~\cite{es_ragas_2025} designed for assessing RAG systems. We compute three indicators, answer relevance, answer correctness, and faithfulness, using the locally hosted \texttt{nomic-embed-text} embedding model and the \texttt{meta-llama/llama-4-scout-17b-16e-instruct} judge served via Groq's low-latency inference backend. This configuration measures grounding and factual alignment without dependency on a commercial grader. To further reduce evaluator bias, the third evaluation repeats the RAGAS computation on eight metrics (LLM Context Precision, Context Recall, Faithfulness, Answer Accuracy, Context Relevance, Response Groundedness, Factual Correctness, and Semantic Similarity) using the open-weight \texttt{Gemma3-27B} model as judge. The results are summarized in Table~\ref{tab:rag_metrics_combined}.

\begin{figure}[H]
  \centering
  \begin{subfigure}{0.45\linewidth}
    \centering
    \includegraphics[width=\linewidth]{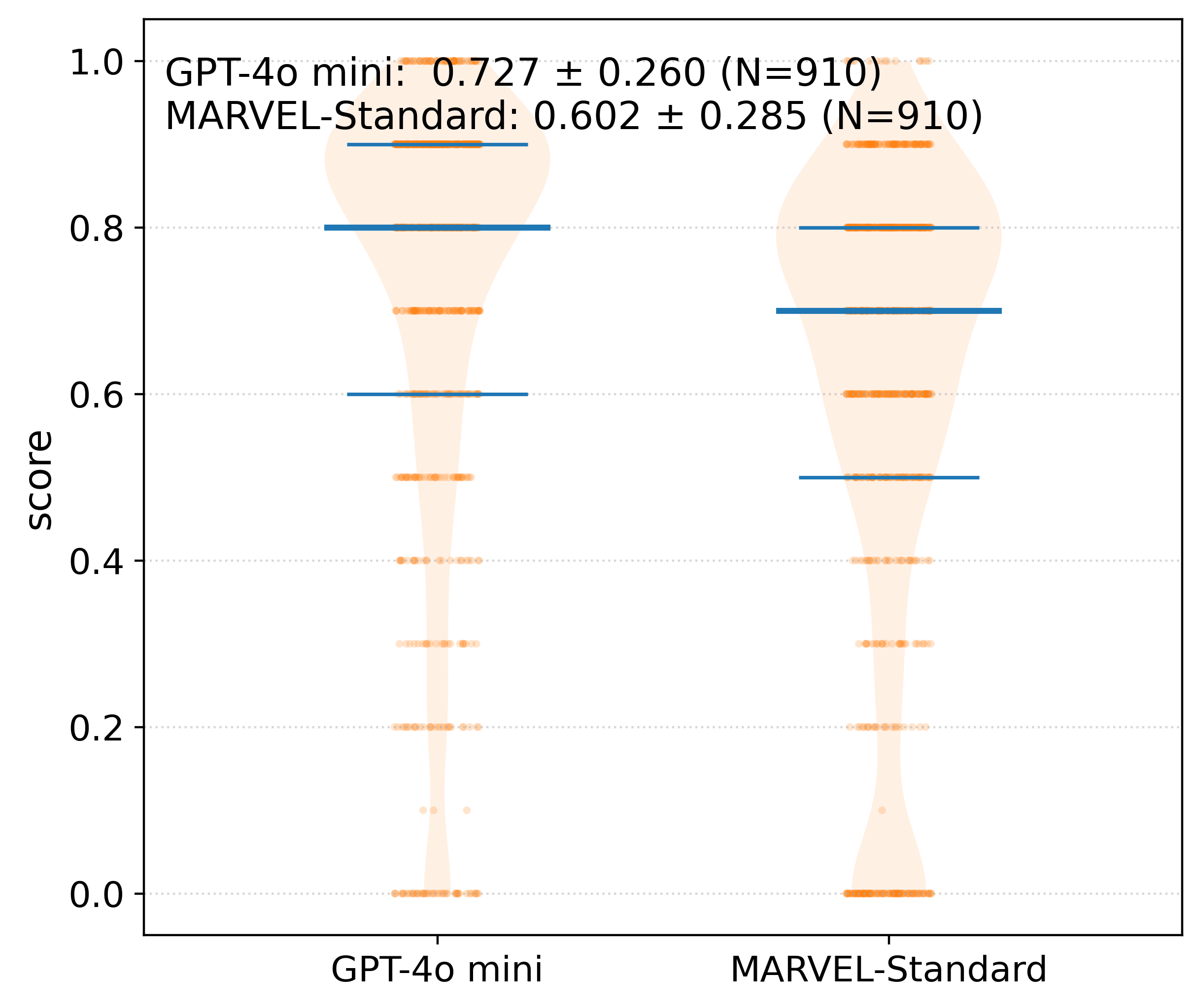}
    \caption{arXiv Data}
    \label{fig:std-arxiv}
  \end{subfigure}
  \begin{subfigure}{0.45\linewidth}
    \centering
    \includegraphics[width=\linewidth]{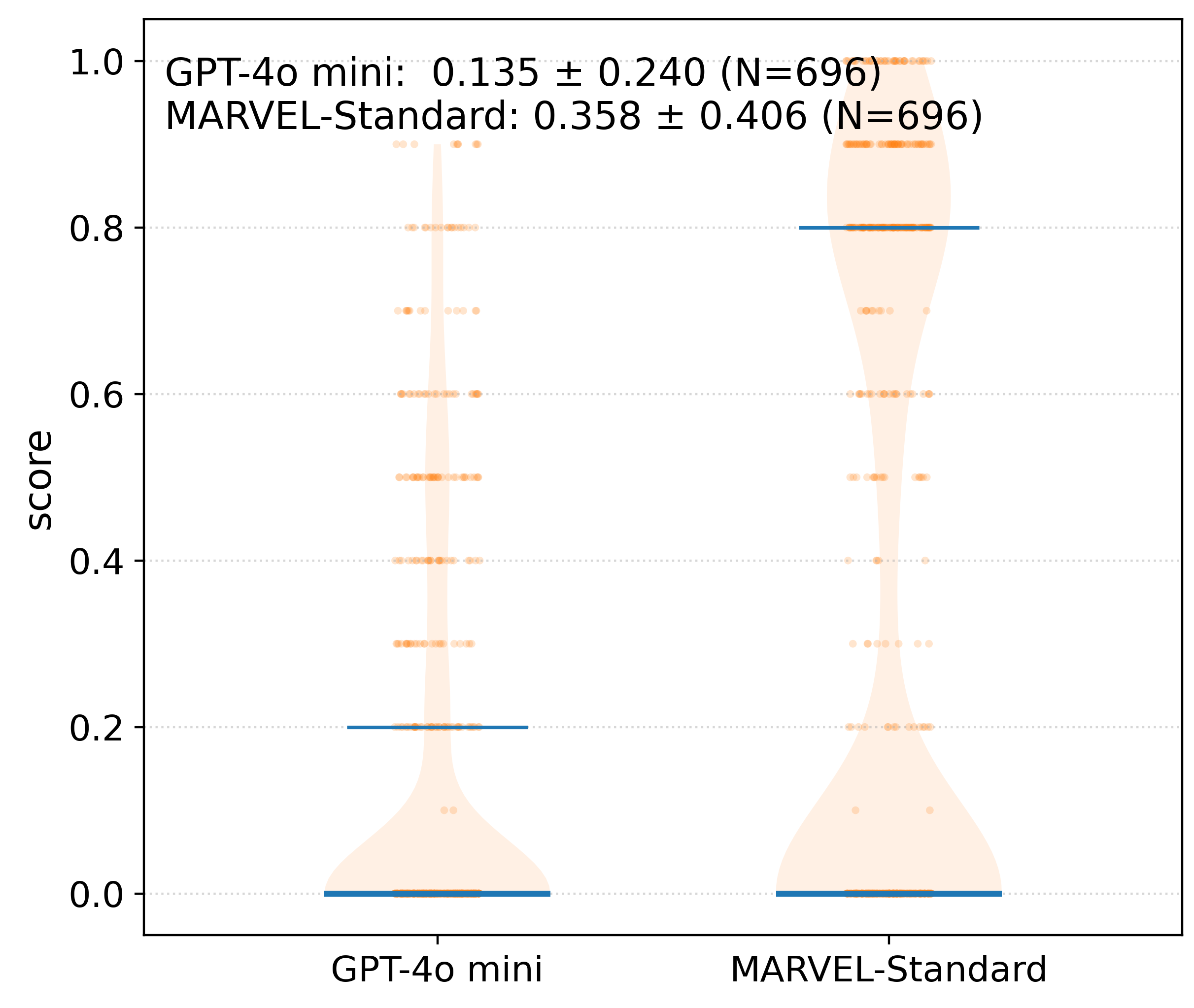}
    \caption{LIGO Logbook Data}
    \label{fig:std-ligo}
  \end{subfigure}
  \caption{Blind A/B judging (GPT-4o mini as judge) on two publicly available surrogate datasets. Violin plots compare GPT-4o mini vs. MARVEL-Standard. On ArXivData, the judge slightly favors GPT-4o mini over MARVEL-Standard. On LogbookData, MARVEL-Standard is preferred, reflecting the benefit of domain retrieval on operational records.}
  \label{fig:violins-standard}
\end{figure}

\begin{figure}[H]
  \centering
  \begin{subfigure}[b]{0.95\textwidth}
    \centering
    \includegraphics[width=\textwidth]{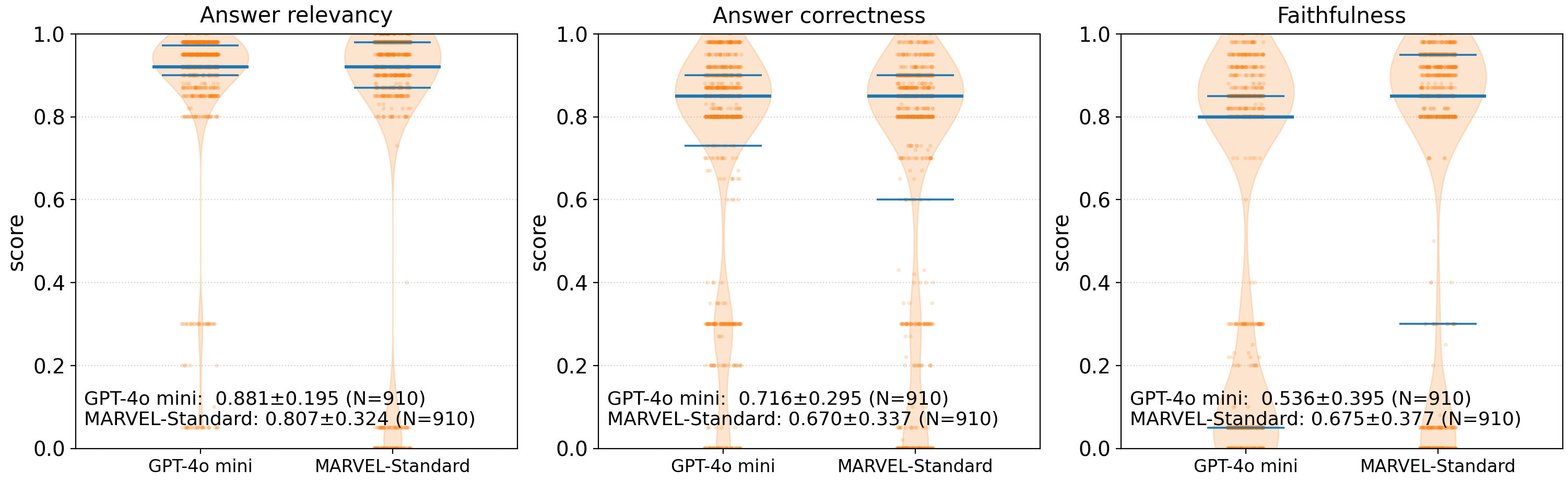}
    \caption{Ten years of arXiv papers containing the keyword "LIGO"}
    \label{fig:ragas-arxiv}
  \end{subfigure}
  \vskip\baselineskip 
  \begin{subfigure}[b]{0.95\textwidth}
    \centering
    \includegraphics[width=\textwidth]{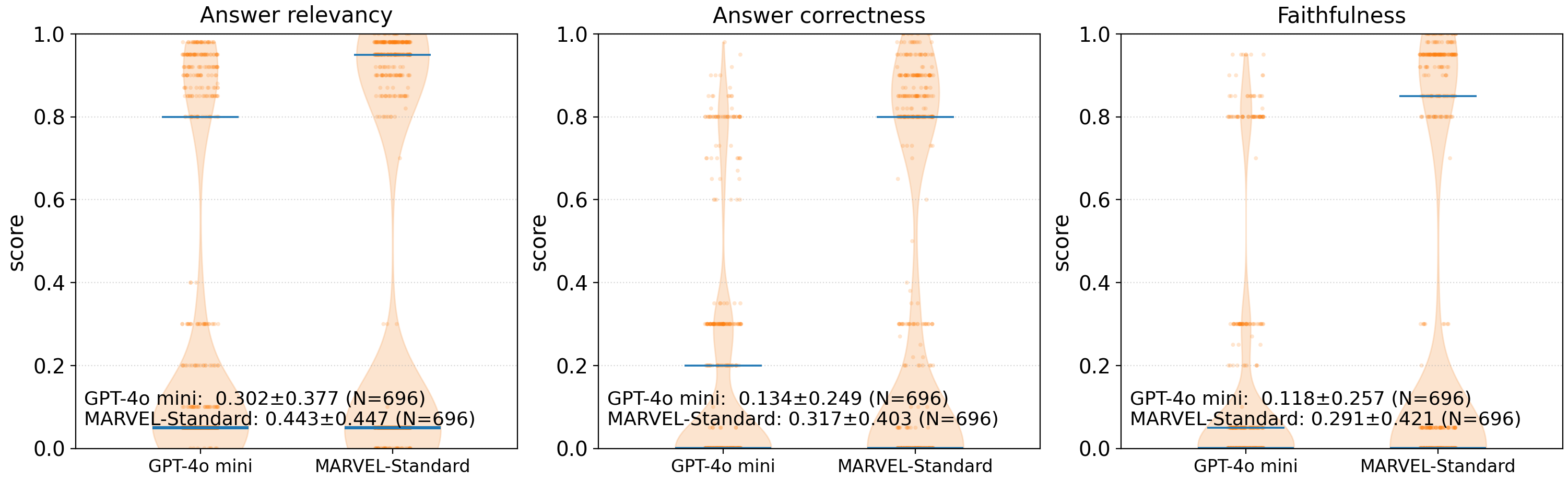}
    \caption{LIGO detector logbook data}
    \label{fig:ragas-logbook}
  \end{subfigure}
  \caption{Evaluation of answer quality for GPT-4o mini baseline and MARVEL-Standard on different datasets. Answers are analyzed and evaluated using open-weight LLM, \texttt{meta-llama/llama-4-scout-17b-16e-instruct}.}
  \label{fig:ragas}
\end{figure}

On the ArXivData dataset, GPT-4o mini and MARVEL-Standard perform almost equally well in the blind A/B tests. The average normalized score given by the GPT-4o mini grader is slightly higher for GPT-4o mini itself (mean $0.73 \pm 0.26$) than for MARVEL-Standard (mean $0.60 \pm 0.29$) (Figure \ref{fig:std-arxiv}). The RAGAS distributions computed on the full ArXivData set (Figure \ref{fig:ragas-arxiv}) show that MARVEL-Standard improves answer faithfulness and accuracy while keeping context relevance and semantic similarity at similar levels.
On the LogbookData dataset, MARVEL-Standard performs clearly better. The blind grader gives mean scores of $0.36 \pm 0.41$ for MARVEL-Standard and $0.14 \pm 0.24$ for GPT-4o mini (Figure \ref{fig:std-ligo}). The RAGAS results on the full LogbookData set (Figure \ref{fig:ragas-logbook}) also show steady improvements in relevance, correctness, and faithfulness.
Table \ref{tab:rag_metrics_combined} reports RAGAS scores for the subset of 168 ArXivData questions and 135 LogbookData questions that were also evaluated with MARVEL-DeepSearch, so that we can compare GPT-4o mini, MARVEL-Standard, and MARVEL-DeepSearch on the same queries. On this common subset, MARVEL-Standard performs better than GPT-4o mini on most metrics. For LogbookData, LLM Context Precision increases from 0.15 to 0.29, Context Recall from 0.29 to 0.36, Faithfulness from 0.16 to 0.33, Answer Accuracy from 0.26 to 0.34, Response Groundedness from 0.21 to 0.35, and Semantic Similarity from 0.35 to 0.43, while Context Relevance stays at 0.93 for both systems. For ArXivData, MARVEL-Standard improves LLM Context Precision (0.36 to 0.56), Context Recall (0.50 to 0.61), and Factual Correctness (0.49 to 0.61), with Context Relevance fixed at 0.85. MARVEL-DeepSearch gives further gains on several metrics for both datasets, especially on recall, groundedness, accuracy, and factual correctness. Although the table uses a smaller subset, the overall pattern is consistent with the full-dataset plots.

\begin{table}[H]
\centering
\resizebox{\textwidth}{!}{%
\renewcommand{\arraystretch}{1.35}
\begin{tabular}{|p{3.8cm}|p{6.8cm}|ccc|ccc|}
\hline
\multirow{2}{*}{\textbf{Metric Name}} &
\multirow{2}{*}{\textbf{Description}} &
\multicolumn{3}{c|}{\textbf{ArXivData}} &
\multicolumn{3}{c|}{\textbf{LogbookData}} \\
\cline{3-8}
& &
\rule{0pt}{2.8ex}\textbf{GPT-4o mini} &
\shortstack{\textbf{MARVEL-}\\\textbf{Standard}} &
\shortstack{\textbf{MARVEL-}\\\textbf{DeepSearch}} &
\rule{0pt}{2.8ex}\textbf{GPT-4o mini} &
\shortstack{\textbf{MARVEL-}\\\textbf{Standard}} &
\shortstack{\textbf{MARVEL-}\\\textbf{DeepSearch}} \\
\hline
Context Relevance & Evaluates whether the retrieved context is relevant to the query.
  & 0.85 & 0.85 & 0.85 & 0.93 & 0.93 & 0.93 \\ \hline
LLM Context Precision & Checks whether the response uses only information present in the provided context.
  & 0.36 & \textbf{0.56} & 0.53 & 0.15 & 0.29 & \textbf{0.32} \\ \hline
Context Recall & Checks whether the response incorporates all relevant information from the context.
  & 0.50 & 0.61 & \textbf{0.66} & 0.29 & 0.36 & \textbf{0.48} \\ \hline
Faithfulness & Evaluates whether the response remains consistent with the provided context.
  & 0.41 & 0.59 & \textbf{0.60} & 0.16 & 0.33 & \textbf{0.37} \\ \hline
Answer Accuracy & Measures the correctness of the response relative to the reference answer.
  & 0.48 & 0.59 & \textbf{0.62} & 0.26 & 0.34 & \textbf{0.42} \\ \hline
Response Groundedness & Checks whether the retrieved context supports the response.
  & 0.47 & 0.61 & \textbf{0.63} & 0.21 & 0.35 & \textbf{0.41} \\ \hline
Factual Correctness & Measures factual correctness relative to the context or facts.
  & 0.49 & 0.61 & \textbf{0.65} & 0.24 & 0.35 & \textbf{0.43} \\ \hline
Semantic Similarity & Measures similarity between the response and the reference answer.
  & 0.60 & 0.66 & \textbf{0.69} & 0.35 & 0.43 & \textbf{0.47} \\ \hline
\end{tabular}%
}
\caption{RAGAS evaluation of response quality across multiple metrics, including contextual precision, recall, faithfulness, and factual grounding. On both ArXivData and LogbookData, the MARVEL-Standard improves over the GPT-4o mini baseline on most metrics, and the MARVEL-DeepSearch variant provides further gains, especially in recall, groundedness, accuracy, and factual correctness. (Evaluation model: google/gemma-3-27b.)}
\label{tab:rag_metrics_combined}
\end{table}

\begin{figure}[H]
  \centering
  \begin{subfigure}{0.48\linewidth}
    \centering
    \includegraphics[width=\linewidth]{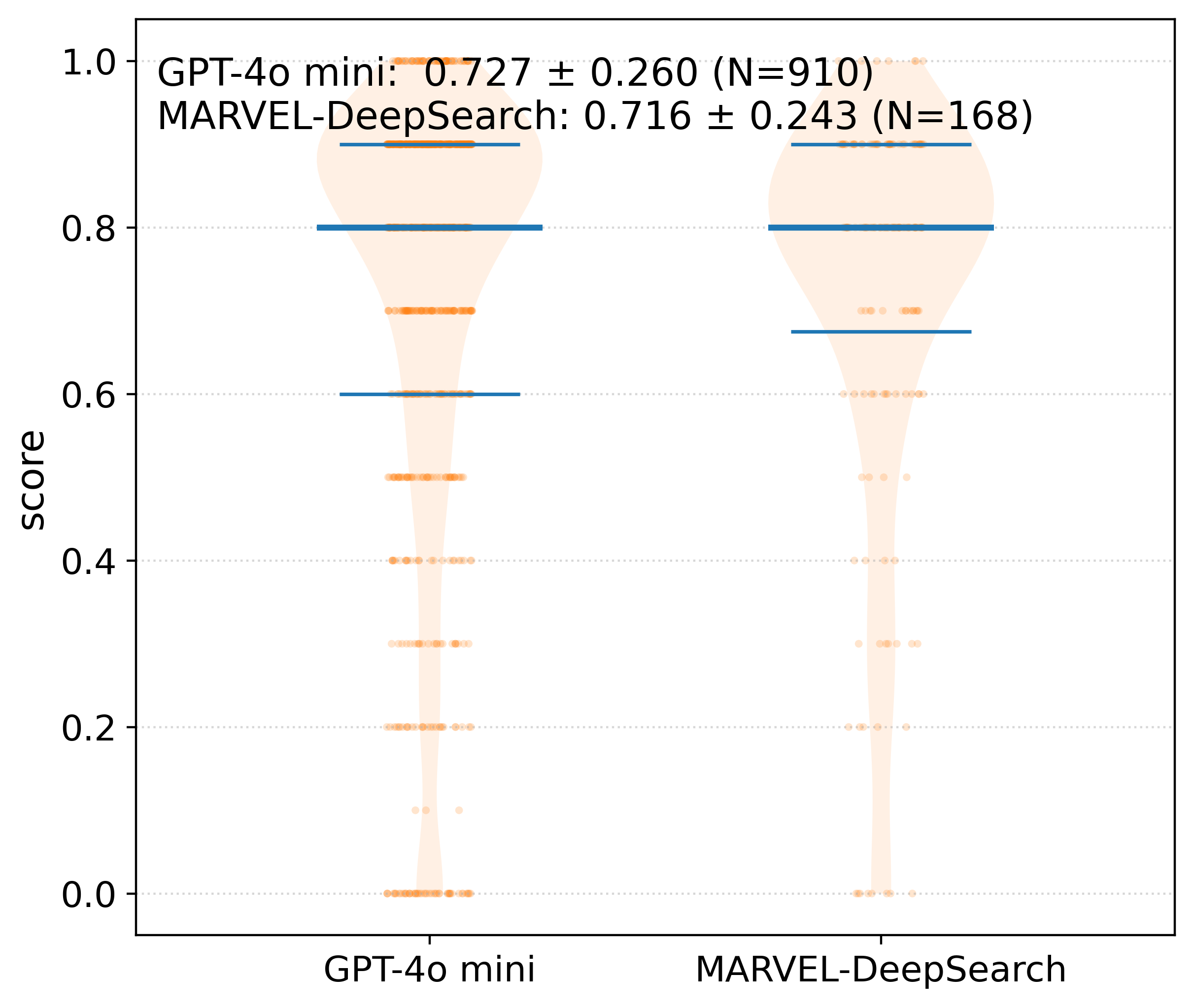}
    \caption{ArXivData}
    \label{fig:deep-arxiv}
  \end{subfigure}\hfill
  \begin{subfigure}{0.48\linewidth}
    \centering
    \includegraphics[width=\linewidth]{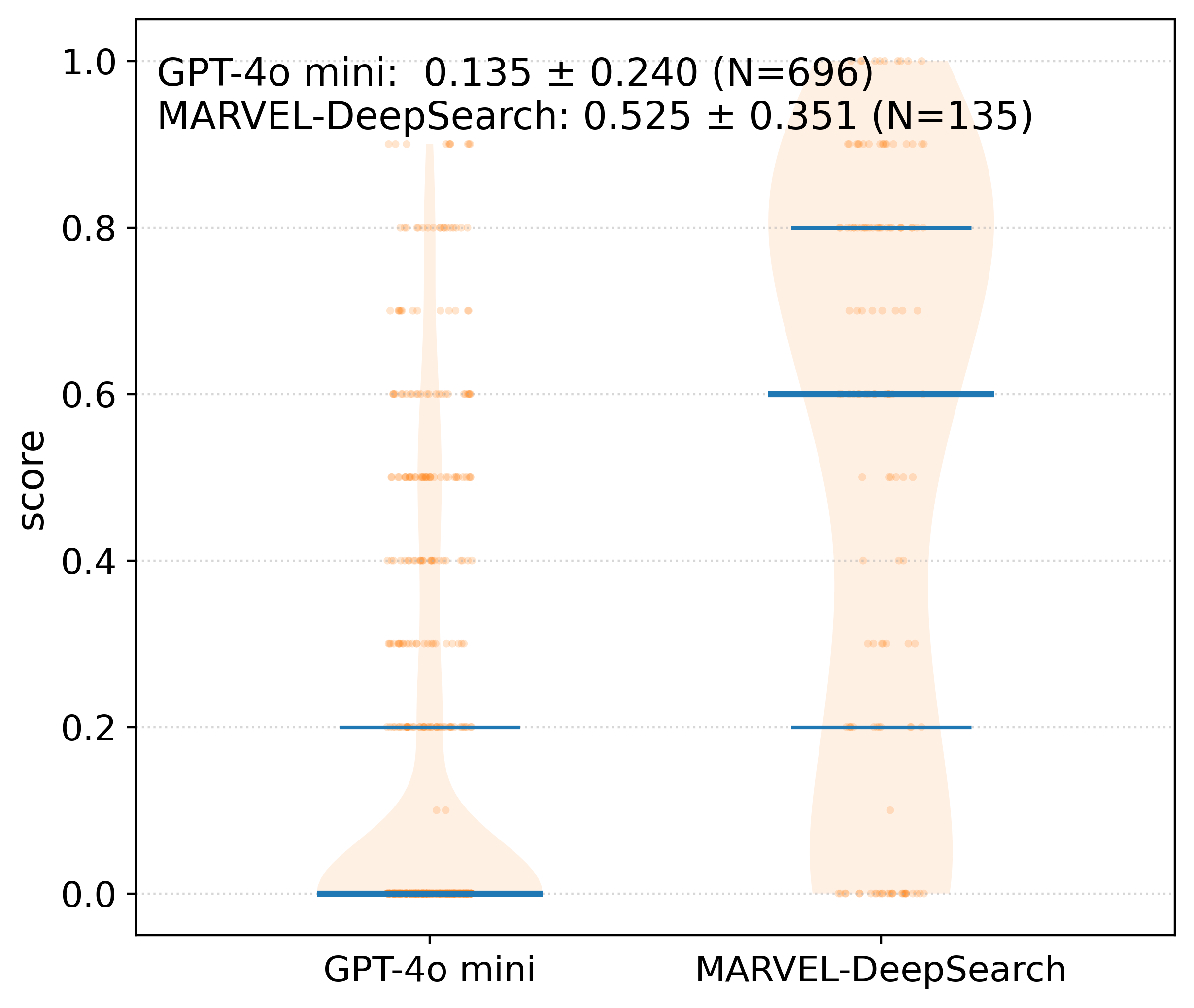}
    \caption{LIGO Logbook Data}
    \label{fig:deep-ligo}
  \end{subfigure}
  \caption{GPT-4o mini vs MARVEL-DeepSearch carried out using Monte Carlo Tree Search with Retrieval Augmented Generation: Violin distributions for arXiv and LIGO Logbook datasets.}
  \label{fig:violins-deepsearch}
\end{figure}

\begin{figure}[H]
  \centering
  \begin{subfigure}[b]{0.95\textwidth}
    \centering
    \includegraphics[width=\textwidth]{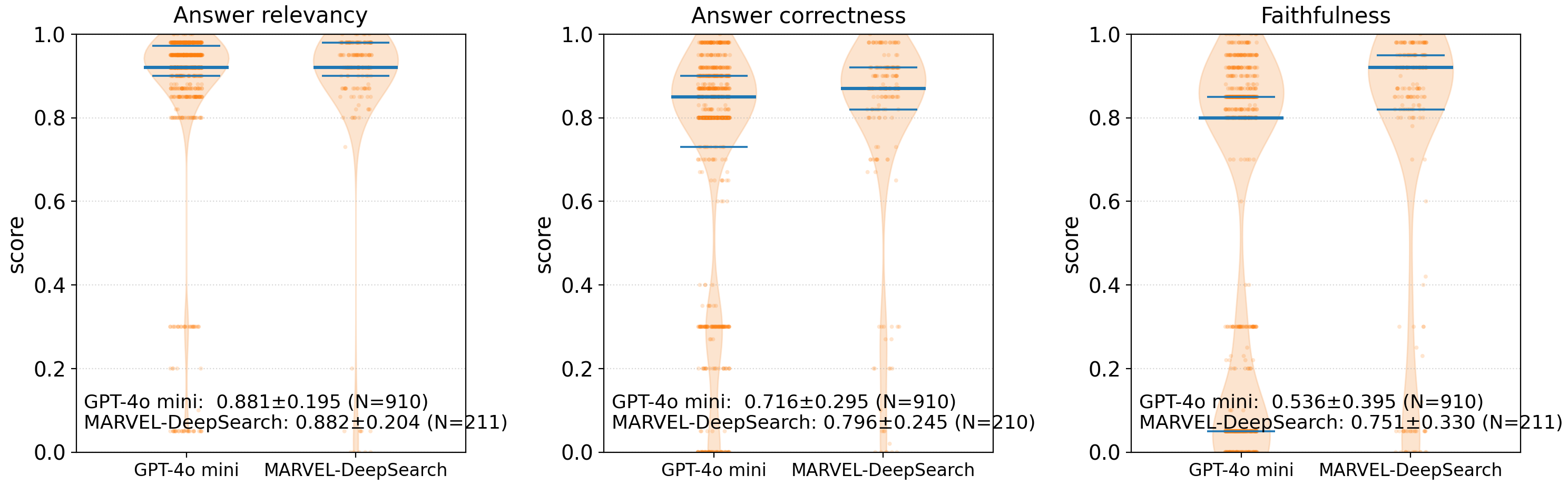}
    \caption{Ten years of arXiv papers related to "LIGO"}
    \label{fig:sub1}
  \end{subfigure}
  \vskip\baselineskip 
  \begin{subfigure}[b]{0.95\textwidth}
    \centering
    \includegraphics[width=\textwidth]{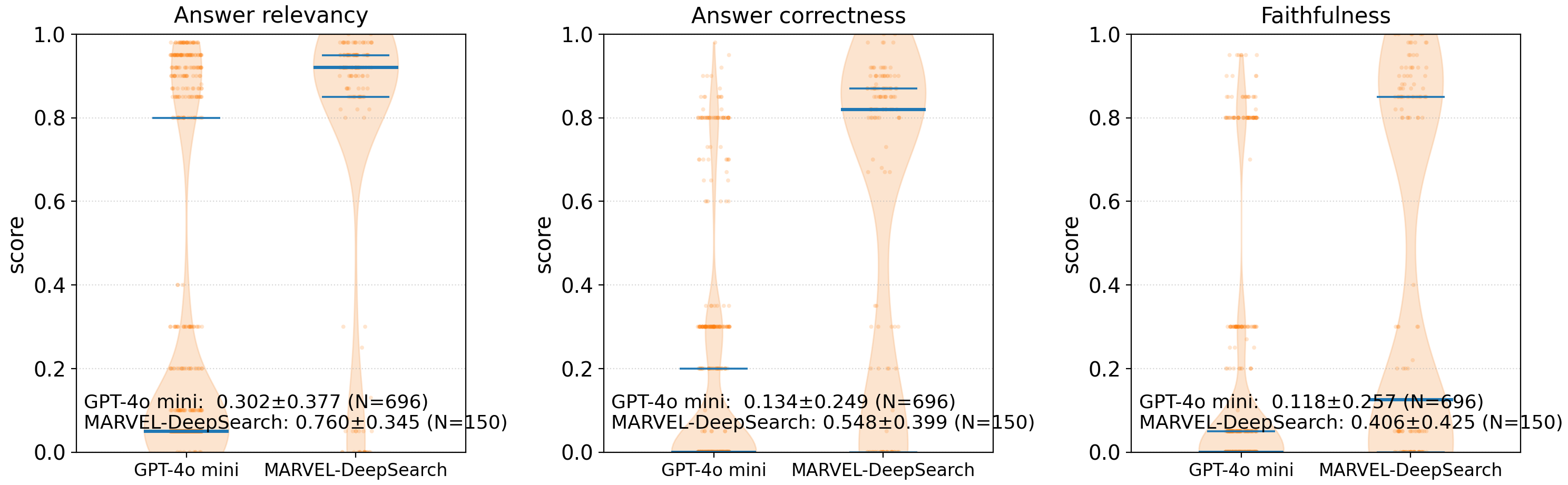}
    \caption{LIGO detector logbook data}
    \label{fig:sub2}
  \end{subfigure}
  \caption{Evaluation of answer quality for GPT-4o mini baseline and MARVEL-DeepSearch on different datasets. Answers are analyzed and evaluated using GPT-4o mini.}
  \label{fig:ragas-deepsearch}
\end{figure}

Because MCTS expansion is computationally heavy, we do not rerun every question. Instead, DeepSearch is evaluated on a representative random sample of about 300 questions taken from both datasets (168 from ArXivData and 135 from LogbookData). Each query is reprocessed under the same retrieval and generation settings, with additional compute allocated for guided exploration via MCTS.
In the blind A/B test, DeepSearch performs on par with GPT-4o mini on ArXivData ($0.73 \pm 0.26$ vs.\ $0.73 \pm 0.24$; Figure \ref{fig:deep-arxiv}) and performs clearly better on LogbookData ($0.52 \pm 0.35$ vs.\ $0.13 \pm 0.26$; Figure \ref{fig:deep-ligo}). The RAGAS evaluation~\cite{es_ragas_2025} shows the same trend. On ArXivData, answer relevance stays comparable ($0.88 \pm 0.22$ vs.\ $0.88 \pm 0.20$), while correctness and faithfulness improve from $0.72 \pm 0.30$ and $0.54 \pm 0.40$ to $0.81 \pm 0.23$ and $0.79 \pm 0.30$, respectively (Figure \ref{fig:sub1}). On LogbookData, DeepSearch shows clear gains across all three metrics: relevance ($0.81 \pm 0.28$ vs.\ $0.30 \pm 0.38$), correctness ($0.57 \pm 0.36$ vs.\ $0.13 \pm 0.25$), and faithfulness ($0.38 \pm 0.41$ vs.\ $0.12 \pm 0.26$) (Figure \ref{fig:sub2}). Although this subset is smaller than the complete evaluation corpus, it captures representative performance trends for both domains. DeepSearch uses the same retriever, ColBERTv2 re-ranker, and open-weight generator as MARVEL-Standard, so the observed improvements arise purely from the adaptive exploration mechanism. Overall, the findings indicate that combining MCTS with RAG leads to better coverage and stronger factual consistency. The improvement is most noticeable for operational records, where the relevant information is usually scattered across many documents.

\section{Conclusions and Outlook}
\label{sec:conclusion}

We tested MARVEL on tasks related to gravitational-wave detectors, running it entirely on open-weight LLMs. Using two publicly available datasets, the MARVEL-Standard setup performed at the level of the GPT-4o mini reference system for literature-oriented queries and did better on queries related to detector operations. We also found that the DeepSearch mode, which allows extra computation during the search, gave a further improvement in answer quality. Making the entire system open source and running on modest hardware lowers the entry barrier for students and early-career researchers. It allows participation from communities that may not have access to commercial platforms due to high licensing or API costs. Developing a large proprietary model such as GPT-4o mini demands large investments, whereas deploying and extending this framework fits easily within the scope of typical research projects and requires much less energy and computing power. As a personalized learning tool, it can be run locally on a laptop using institution-specific or user-chosen data to produce explanations with citations to the chosen sources.

A possible direction for future work is to let the system adjust its behaviour based on user feedback. One way would be to use reinforcement learning as it can guide the model toward expert preferences~\cite{li2023reinforcement}. 
Going beyond retrieval-based reasoning, we intend to add capabilities that enable MARVEL to interact with real or simulated environments. As part of standard safety checks, the agent could be tasked with inspecting data, performing actions, and recording resulting state changes.
 If successful, such an agent can support activities at detection sites. Overall, the MARVEL framework provides a foundation for applying open-weight LLMs to domain-specific scientific tasks, on which more capable systems can be developed in the future.

\section*{Code and Data Availability}
\label{sec:availability}

A public demo of MARVEL is available at \url{https://ligogpt.mit.edu/marvel}.
The source code, configuration files, and QA datasets used for evaluation are released as MARVEL v1.0.0 on GitHub:
\url{https://github.com/Nikhil-Mukund/marvel/releases/tag/v1.0.0}.
For long-term preservation and citation, the same release is archived on Zenodo~\cite{marvel_zenodo_18156827}:
\url{https://zenodo.org/records/18156827}.

\section*{Acknowledgments}

NM, FZ, LB, and EK acknowledge support from the U.S. National Science Foundation (NSF)
under awards PHY-1764464 and PHY-2309200. NM and LB acknowledge support from NSF under
Cooperative Agreement PHY-2019786 (The NSF AI Institute for Artificial Intelligence and
Fundamental Interactions, \url{http://iaifi.org/}) and from MathWorks, Inc. EK acknowledges
support from NSF under PHY-2117997 (The NSF Institute on Accelerated AI Algorithms for
Data Driven Discovery - A3D3, \url{http://a3d3.ai/}). NM, LB, and EK thank the MIT Office
of Research Computing and Data (ORCD) for the seed grant and Jonathan Murray for assistance
with using the MIT Engage cluster and H200 GPUs for the analysis. We thank Adam Zacharia
Anil for his help in setting up the MARVEL website and Tiago Fernandes for providing initial
feedback. We also thank Frederik Donovan and Paul Hsi for technical support in setting up
and running MARVEL. The authors are grateful for computational resources provided by the
LIGO Laboratory and supported by the National Science Foundation Grants PHY-0757058 and
PHY-0823459. This material is based upon work supported by NSF's LIGO Laboratory which
is a major facility fully funded by the National Science Foundation.

\printbibliography

\end{document}